\documentclass[prd,twocolumn,nofootinbib]{revtex4}
\usepackage{cancel}
\usepackage{graphicx}
\usepackage{epsfig}
\usepackage{amsmath,amsfonts,amssymb}
\usepackage{t1enc}

\begin{document}

\preprint{CAFPE-11-xxx}
\preprint{UGFT-11-xxx}  

\title{Four tops and the $t\bar{t}$ forward-backward asymmetry}

\author{J. A. Aguilar-Saavedra}
\author{J. Santiago}
\affiliation{
CAFPE and Departamento de F\'{\i}sica Te\'orica y del Cosmos,
\\
Universidad de Granada, E-18071 Granada, Spain
}


\begin{abstract}
New colour octet vectors below the TeV scale could explain the anomalous
$t\bar{t}$ forward-backward 
asymmetry observed at the Tevatron experiments, while being consistent
with the current LHC data. These models  
generally lead to four-top final states at the LHC at observable
levels. We compute the four-top production cross section at the LHC 
in a model with a massive colour octet vector as a function its mass,
its width and its coupling to the top quark.
Octet masses in the 
vicinity of the $t\bar{t}$ threshold are generally excluded by present
limits on the production of same-sign dileptons and trileptons. 
Masses above 650 GeV are allowed, quite independently of the
couplings, but they can be probed with the luminosity of $5$
  fb$^{-1}$ already
collected at the LHC, up to around 800 GeV. 
The four-top production
cross section is increased by a factor $\sim 2$ with $\sqrt{s}=8$ TeV
and by up to almost two orders of magnitude with $\sqrt{s}=14$ TeV, thus
greatly increasing the reach for massive gluons after the LHC energy upgrade.
\end{abstract}

\maketitle


Due to its large mass, the top quark is expected to play a relevant
role in the discovery of new physics beyond the Standard Model (SM). The
first hint of such new physics could be already available in the form
of the anomalously large $t\bar{t}$ forward-backward (FB) asymmetry
observed at both Tevatron experiments~\cite{arXiv:1101.0034,arXiv:1107.4995}.
The fact that neither the Tevatron nor the Large Hadron Collider (LHC)
have observed any other anomaly in top or jet physics sets strong
constraints on possible explanations in terms of new
physics~\cite{Hewett:2011wz,arXiv:1105.4606,arXiv:1107.0841}. One of the few
surviving explanations, compatible with the present measurements of the $t \bar t$ invariant mass spectrum~\cite{tailCMS} and the charge asymmetry at the LHC~\cite{afbCMS2,afbATLAS}, is a relatively light colour octet vector boson
(called here `gluon' for brevity) with mass $M \lesssim 1$ TeV and
with suppressed axial-vector couplings to the light 
quarks and sizeable axial-vector couplings to top quarks~\cite{Barcelo:2011fw,Haisch:2011up,arXiv:1106.4054,arXiv:1107.0978,arXiv:1107.1473,
arXiv:1107.2120,arXiv:1109.0648,arXiv:1110.3796,Barcelo:2011wu}. 
The axial coupling ensures cancelation
of the interference terms between the SM  and new physics
contribution~\cite{Ferrario:2008wm,arXiv:1010.6304,arXiv:1103.2765} 
in symmetric observables
while preserving the contribution to the asymmetry. Masses around 1
TeV require large couplings to the top quark and a large gluon width,
usually with extra decay 
channels~\cite{arXiv:1106.4054}. Masses close to the $t\bar{t}$
threshold can easily hide in the large SM $t\bar{t}$ background,
although they may also need extra decay channels to be 
invisible~\cite{arXiv:1107.0978}. Masses lighter than the $t\bar{t}$
threshold can do with smaller couplings and are essentially invisible
in symmetric 
observables~\cite{arXiv:1107.2120}.

In this paper we consider an alternative, yet unexplored probe of these models. The
massive gluon is a colour octect vector resonance, thus its couplings
to SM gluons are fixed by gauge invariance. Due to the relatively low
masses relevant for the FB asymmetry, pair production of
such objects with subsequent decay in two top pairs can receive a
fairly large cross section, which is further increased by non-resonant
contributions and by single gluon resonant production, 
especially if the coupling of the new gluon 
to the top quark is
sizeable. (See~\cite{Dicus:1994sw,Lillie:2007hd,Dobrescu:2007yp,Perelstein:2011ez,Cacciapaglia:2011kz}   
for preliminary studies of colour octet pair production at hadron
colliders followed by top decays.) 
Four-top final states have a very small
background in the SM but are difficult to reconstruct completely (see
for instance section 12 of~\cite{Brooijmans:2010tn} 
and~\cite{Servant:2010zz}). Here we
show that simpler searches, based on production of same-sign dileptons and trileptons, are
enough to probe and constrain models of light gluons as an
explanation of the $t\bar{t}$ asymmetry. Specifically, to estimate the
present limits on four-top production at the LHC we use (i) a
supersymmetry-motivated search~\cite{arXiv:1110.2640}; (ii) a search
for fourth generation $b'$ quarks~\cite{trilep}, both performed by the
CMS Collaboration. The present analysis is of course relevant to any
  model with color octect vector resonances that couple strongly to
  the top quark and not only to the ones attempting to explain the top
  FB asymmetry.

\begin{figure}[htb]
\begin{center}
\epsfig{file=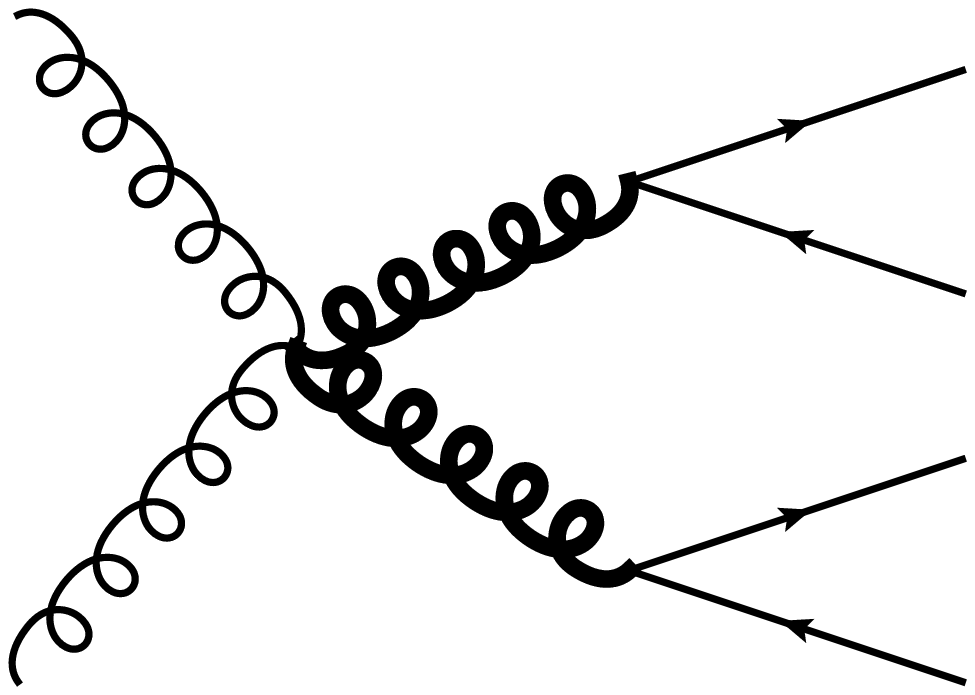,height=2.5cm,clip=} 
\hspace{22pt}\epsfig{file=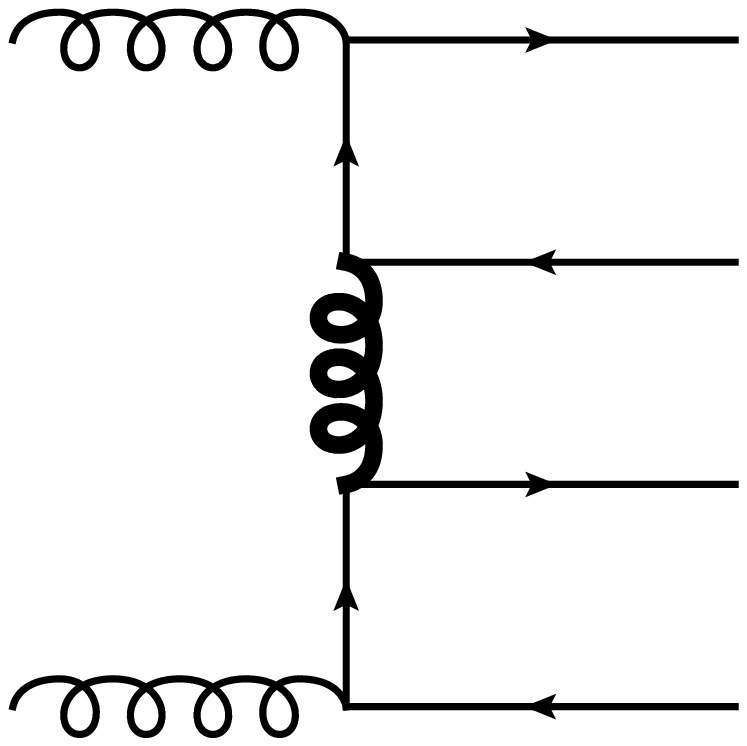,height=2.5cm,clip=} 
\caption{Sample diagrams for resonant (left) and non-resonant (right)
  contribution to four-top production in the presence of new heavy
  gluons. The thick line  corresponds to the massive gluon.}
\label{fig:diagrams}
\end{center}
\end{figure}

Let us consider a new massive gluon $G$. Its couplings to SM gluons
are fixed by gauge invariance 
whereas the ones to fermions $g_i^{V,A}$ are in principle free. The
relevant Lagrangian is 
\begin{eqnarray}
\mathcal{L}^G & = & -\frac{1}{2} D_\mu G^a_\nu \Big(D^\mu G^{a\,\nu}
-D^\nu G^{a\,\mu}\Big)+\frac{1}{2}M^2 G_\mu^a G^{a\,\mu} \notag \\
& & + \bar{\psi}_i \gamma^\mu G^a_\mu \frac{\lambda^a}{2} 
\Big[g_i^V + g_i^A \gamma_5\Big] \psi_i,
\end{eqnarray}
where $i$ is a flavor index, $\lambda^a$ are the Gell-Mann matrices and
\[
D_\mu G^a_\nu 
\equiv \partial_\mu G^a_\nu  
+ g_s f^{abc} g_\mu^b G_\nu^c
\]
is the SM covariant derivative, with 
 $a=1,\ldots,8$, $g_\mu^a$ the SM gluons, $f^{abc}$ the SU(3)
structure constants and $g_s$ the strong coupling
constant. 
In order to contribute to the $t \bar t$ asymmetry, $G$ must have
non-vanishing couplings to the top and first generation quarks, being 
the FB asymmetry in $q \bar q \to t \bar t$ proportional to the product $g_q^A
g_t^A$, with $q=u,d$. Searches for dijet resonances typically
constrain $g^A_q$ (as well as the vector coupling) to be relatively
small, the precise bound depending on the gluon mass $M$ (see for
  instance~\cite{Haisch:2011up,arXiv:1107.2120}). This implies
that the axial coupling to the top must be of order unity or even
larger, in order to generate a sizeable asymmetry. 
For gluon masses above the $t \bar t$ threshold, $M \geq 2 m_t$, pair
production of massive gluons followed by decays into top pairs is a
large source of four-top final states, see Fig.~\ref{fig:diagrams}
(left). Non-resonant diagrams such as the one depicted in the right
panel, in which the new gluons are not produced on-shell, are also
important for larger values of the gluon coupling to
the top quark, and dominate both below the $t \bar t$ threshold and at
large gluon masses.

Given the fact that $g_q^{A,V} \ll g_t^A$, the cross section for
four-top final states is essentially independent of the coupling to
light quarks. 
For definiteness, we take a purely axial coupling $g_q \equiv g_q^A =
0.2$ to light quarks, which is around the upper limit for a wide range
of heavy gluon masses~\cite{arXiv:1107.2120}, and a right-handed one
$g_t/2 \equiv g_t^A = g_t^V$ to the top quark, as preferred by $B$
physics constraints~\cite{arXiv:1101.5203}. (Setting $g_q$ to zero the
four-top cross section found is nearly identical in all mass range,
except for a slightly steeper rise at the $M \sim 2 m_t$ threshold.)
The coupling to the second generation is also constrained to be small
by dijet production and has even a smaller effect on our results. For
simplicity, it is set to zero. On the other hand, the four-top cross
section depends on the gluon mass and its coupling to the top quark.  
In case that additional new particles exist, the four-top cross
section near and above the $t \bar t$ threshold also depends on the partial width for gluon decays into these
new particles. 

In order to test the sensitivity of existing analyses to four-top production, 
we have implemented our model in {\tt MADGRAPH 5}~\cite{arXiv:1106.0522} using
{\tt FeynRules}~\cite{arXiv:0806.4194}. The matrix element generated
by {\tt MADGRAPH} has been implemented in {\tt
  Protos}~\cite{arXiv:0803.3810} for an efficient exploration of the
model parameter space and computation of four-top production cross
sections. 
 We have generated events for different
configurations of gluon masses and couplings for $pp$ collisions at a 
centre of mass (CM) 
energy  $\sqrt{s}=7$ TeV
and passed them through
{\tt PYTHIA}~\cite{hep-ph/0603175} and {\tt PGS4}~\cite{PGS4}. All our
simulations are performed at leading order. 
For the same-sign dilepton final state we have
applied the selection and kinematical cuts in
Ref.~\cite{arXiv:1110.2640} and found that the analysis most sensitive
to four-top production is the one requiring
\begin{itemize}
\item two same-sign leptons $\ell^\pm \ell^\pm$, $\ell=e,\mu$ with pseudo-rapidity $|\eta|<2.4$. Electrons must have transverse momentum $p_T>10$ GeV and for muons
 $p_T>5$ GeV is required.
\item $H_T\geq 400$ GeV, where $H_T$ is the scalar sum of the $p_T$ of all
jets. Only those with $p_T>40$ GeV and $|\eta|<2.5$ are considered here.
\item Missing energy $\cancel{E}_T\geq 50$ GeV.
\end{itemize}
The global efficiency of these cuts for our four-top signal, including the same-sign dilepton branching ratio, is approximately of $2\%$ for a wide range of heavy gluon masses. (Requiring $H_T\geq 200$ GeV and $\cancel{E}_T\geq 120$
GeV results in an
efficiency only slightly smaller.) With this selection, the CMS Collaboration measures 7 events with an integrated luminosity $L=0.98~\text{fb}^{-1}$, for a SM background prediction of $5.3 \pm 2.4$~\cite{arXiv:1110.2640}. For the trilepton channel we ask for
\begin{itemize}
\item three leptons $\ell=e,\mu$ with $p_T>20$ GeV and $|\eta|<2.4$;
  same-flavour opposite-charge pairs are required to be outside a
  window $|M_Z-m_{ll}| < 10$ GeV ($m_{ll}$ is the invariant mass of
  the two leptons).
\item Two jets with $p_T > 25$ GeV and $|\eta|<2.4$, at least one $b$-tagged.
\item The scalar sum $H_T + \sum_{\ell} p_T^\ell + \cancel{E}_T\geq 50$ must be larger than 500 GeV.
\end{itemize} 
With such cuts the efficiency for our four-top signal is of 0.6\%. With this selection, the CMS Collaboration measures one event with a luminosity $L= 1.16~\text{fb}^{-1}$, for an expected SM background of $0.16\pm 0.09$~\cite{trilep}.

Upper bounds on four-top production can be obtained from either of these channels, as well as from their combination, using the modified frequentist likelihood method~\cite{Junk:1999kv,Read:2000ru}. These limits are evaluated using 
$10^6$ pseudo-experiments of the expected signal and background samples. 
Statistical uncertainty effects are implemented assuming Gaussian 
distributions~\cite{Junk:1999kv}. The obtained $95\%$ CL bound on four-top production are
\begin{align}
& \sigma_{4t} \leq 0.50\mbox{ pb} && (2l) \,, \notag \\
& \sigma_{4t} \leq 0.70\mbox{ pb} && (3l) \,, \notag \\
& \sigma_{4t} \leq 0.36\mbox{ pb} && \mbox{(combined)} \,.
\end{align}

\begin{figure}[htb]
\begin{center}
\epsfig{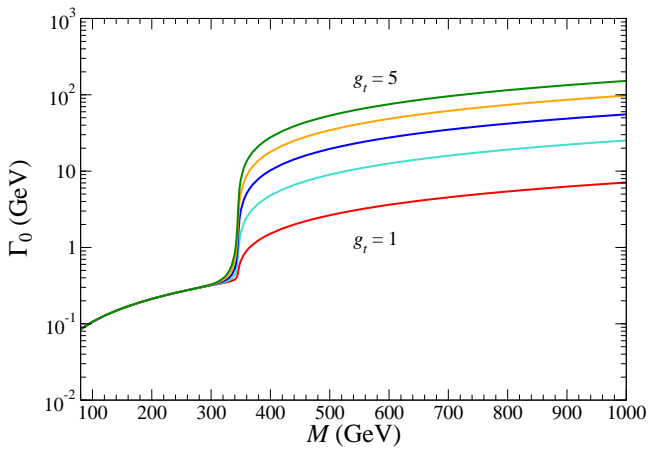} 
\caption{Partial width of the heavy gluon to SM final states. The five lines, from bottom to top, correspond to $g_t=1,2,3,4,5$.}
\label{fig:G0}
\end{center}
\end{figure}

As we have mentioned, the four-top cross section crucially depends on
whether the new gluon can decay to additional non-SM particles. Thus,
a detailed discussion of the heavy gluon width is compulsory. 
Let us denote by $\Gamma_0$ the partial width of the gluon to SM
particles. Below the $M\sim 2 m_t$ threshold, $\Gamma_0$ receives the
largest contribution from decays 
$G \to u \bar u,d \bar d$, with a smaller one from four-body decays $G
\to W^+ b W^- \bar b$. (At any rate, for masses $M \leq 320$ GeV the
four-top cross section is practically independent of $\Gamma$, as we
will explicitly see below.) Above this threshold, $\Gamma_0$ is
largely dominated by on-shell decays to $t \bar t$. This is clearly
seen in Fig.~\ref{fig:G0}, in which we plot $\Gamma_0$ as a function of
$M$, for five values of the coupling to the top quark $g_t=1,2,3,4,5$.

\begin{figure}[htb]
\begin{center}
\epsfig{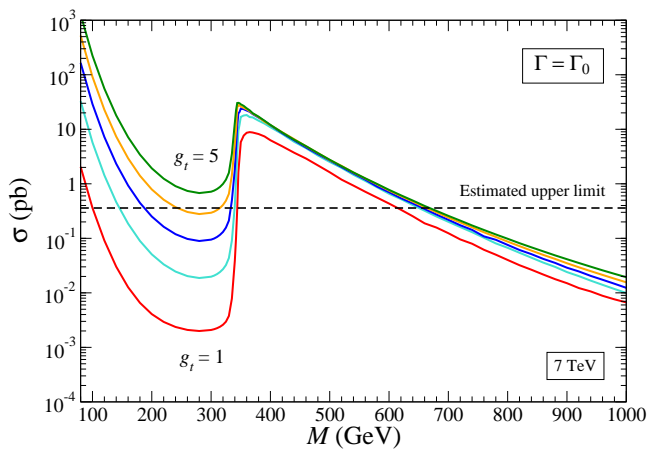} 
\caption{Four-top cross section for $\Gamma=\Gamma_0$ (decays to SM
  particles only) for the LHC with $\sqrt{s}=7$ TeV. 
The five lines, from bottom to top, correspond to $g_t=1,2,3,4,5$.}
\label{fig:xsec-G0}
\end{center}
\end{figure}

We consider in first place models in which the new gluon only decays to SM particles, that is, $\Gamma=\Gamma_0$. The four-top cross section is shown in Fig.~\ref{fig:xsec-G0} for $g_t=1,2,3,4,5$.
Clearly, for $M \geq 2 m_t$ the four-top cross section receives a
boost from diagrams corresponding to on-shell production of two gluons 
with subsequent decay to $t \bar t$. Still, for these masses the
contribution of non-resonant diagrams (and diagrams with a single
on-shell gluon production) is important, as it can be found
out by the different cross sections for the several $g_t$ values
considered. (Clearly, the cross section for on-shell production of two
gluons with subsequent decay to $t \bar t$ is independent of $g_t$, as
long as $G \to t \bar t$ is the dominant decay channel.) From this
plot we can learn that models with gluon masses $M = 350-650$ GeV are
quite generally excluded, unless there is an extra enhancement of the
width by decay to non-SM particles. As soon as the limits on four-top
production at LHC get more stringent, with dedicated analyses and the
use of the full 5 fb$^{-1}$ dataset, larger masses will be
excluded. For example, an upper limit $\sigma_{4t} < 0.1$ pb (slightly better than a naive $1/\sqrt{L}$ rescaling) seems likely, especially bearing in mind the possibility of combination with the semileptonic channel. Such limit will allow to probe gluon masses up to $M \sim 800$ GeV. 

\begin{figure}[htb]
\begin{center}
\epsfig{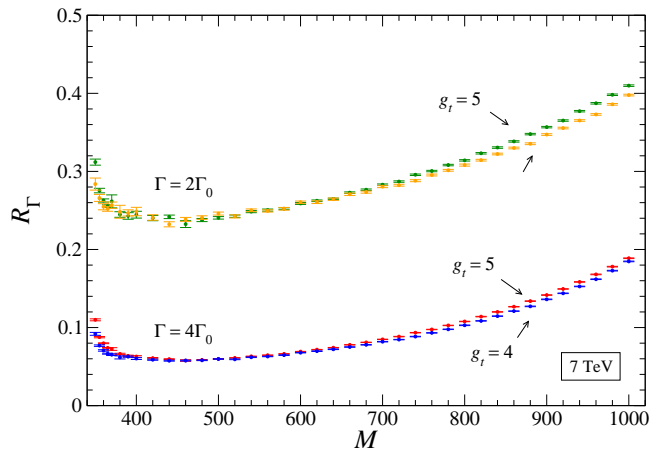} 
\caption{Ratio $R_\Gamma$ in Eq.~(\ref{ec:R}) between cross sections for different values of the gluon total width $\Gamma$, for $M \geq 2 m_t$. The error bars represent the Monte Carlo uncertainty.}
\label{fig:Rhigh}
\end{center}
\end{figure}

Realistic models explaining the FB asymmetry with a new gluon above the $t \bar t$ threshold often require new particles to enhance the gluon width, $\Gamma > \Gamma_0$, so as to make the resonance invisible in the $t \bar t$ invariant mass spectrum~\cite{arXiv:1106.4054,arXiv:1107.2120}. In this case, the four top cross section decreases 
by a factor $R \sim (\Gamma_0/\Gamma)^2$, but not exactly equal to this ratio of widths because of the contributions from diagrams with non-resonant $G$ exchange. We plot in Fig.~\ref{fig:Rhigh} the ratio of cross sections for different gluon total widths,
\begin{equation}
R_\Gamma = \frac{\sigma_{4t}|_\Gamma}{\sigma_{4t}|_{\Gamma_0}} \,,
\label{ec:R}
\end{equation}
for $\Gamma=2\Gamma_0,4\Gamma_0$. We only consider $g_t=4,5$, since these heavy gluon masses require a large top coupling to generate the FB asymmetry.
We observe that this ratio deviates from $(\Gamma_0/\Gamma)^2$ both at
threshold and at high $M$, due precisely to the sizeable contributions
from non-doubly-resonant diagrams. 
Note also that in our simulations we have considered the fixed width
approximation. For very large widths its full energy dependence must
be taken into 
account, which reduces the suppression with respect to the one
depicted in Fig.~\ref{fig:Rhigh}~\cite{arXiv:1106.4054}. Thus,
although the enlarged gluon width required in 
realistic models of the FB asymmetry with gluon masses above threshold
tend to reduce the constraints, they do not remove them
completely. Furthermore, we will see below that once the LHC energy is
upgraded, the dramatic increase in the production cross section will be
enough to impose stringent constraints even with enlarged widths.

\begin{figure}[htb]
\begin{center}
\epsfig{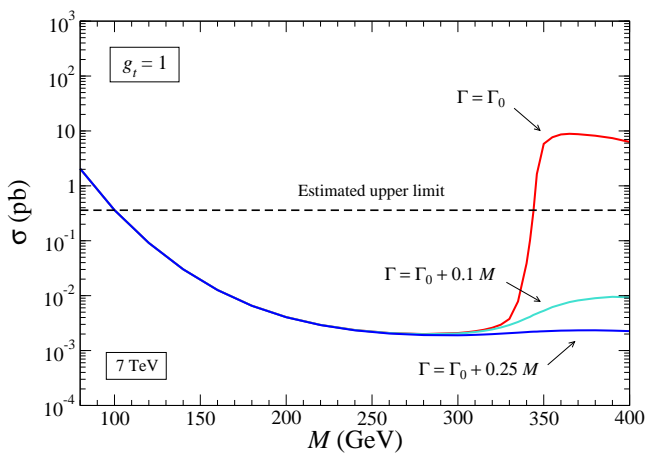} 
\caption{Four-top cross section for new gluons below and slightly above the $t \bar t$ threshold, for $g_t=1$, with and without an extra width enhancement.}
\label{fig:xsec-Gex}
\end{center}
\end{figure}

Models with new gluons of masses $M \sim 300$ GeV under the $t \bar t$ threshold can generate sizeable asymmetries with $g_t$ of order unity~\cite{arXiv:1107.2120}. In this case, four-top production is well below the present and foreseable limits. Still, one may consider a width enhancement from decay to particles lighter than the top quark~\cite{arXiv:1110.3796}. We show in Fig.~\ref{fig:xsec-Gex} the four-top cross section in this case, for width enhancements $\Gamma=\Gamma_0+0.1M$ and $\Gamma=\Gamma_0+0.25M$. In both cases the cross section for masses $M \leq 300$ GeV is unchanged by the extra width, so models with very light colour octets~\cite{arXiv:1109.0648} may already be compromised by limits on four-top production. On the other hand, four-top production close to threshold is largely suppressed, a fact which is expected since the extra width $0.1M$, $0.25M$ to non-SM states is orders of magnitude larger than $\Gamma(G \to t \bar t)$, see Fig.~\ref{fig:G0}. Besides, achieving such a width enhancement may not be natural and/or may require too large couplings to the new particles. At any rate, an extra gluon width may hide the four-top signal but gives rise to other new final states from the decay of the heavy gluons, which have to be considered when discussing the viability of any model.

\begin{figure}[htb]
\begin{center}
\epsfig{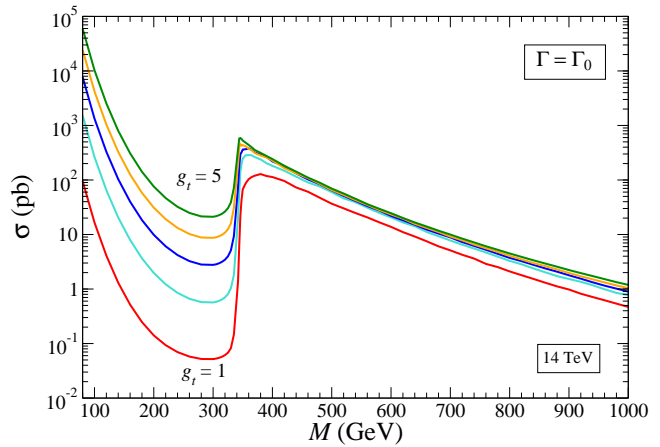} 
\caption{
Four-top cross section for $\Gamma=\Gamma_0$ (decays to SM particles
only) for the LHC with $\sqrt{s}=14$ TeV. 
The five lines, from bottom to top, correspond to
$g_t=1,2,3,4,5$.} 
\label{fig:4tat14}
\end{center}
\end{figure}

Let us now consider the effect of the foreseen LHC energy
  upgrade. If the CM energy is increased to $8$ TeV we
  obtain a factor of $2-2.3$ increase in the four-top production
  cross section, thus partially compensating the supression due to
  an enlarged gluon width (see Fig.~\ref{fig:Rhigh}). A much more
  dramatic increase of the signal cross section is obtained for
  $\sqrt{s}=14$ TeV, as we show in Fig.~\ref{fig:4tat14}. The four-top
production cross section in enhanced by one to almost two orders of
magnitude, depending on the gluon mass, with respect to the one at
$\sqrt{s}=7$ TeV. Although the backgrounds also grow at this energy we
can anticipate a very good sensitivity to four-top production. For
example, the lowest point in Fig.~\ref{fig:4tat14} has a cross section
of 52 fb while an estimated 5 $\sigma$ observation limit of 45 fb is
expected with 100 fb$^{-1}$ of integrated luminosity~\cite{Lillie:2007hd}.
Moreover, the production cross section
at the $t\bar{t}$ threshold is almost four orders of magnitude larger
than the expected observation limit. Thus, even with a strong suppresion
due to an enlarged width, models with a light gluon below the TeV scale are
expected to be probed at the LHC.

In summary, in this paper we have considered four-top production in
models explaining the Tevatron $t \bar t$ asymmetry with new `light'
gluons. Pair production of these particles followed by decays into top
pairs is a new, potentially large, source of four-top final states.  
In order to cover all the relevant parameter space, we have studied
the four-top production cross section as a function of the gluon mass
and its coupling to the top. We have also considered some examples of
scenarios where the heavy gluon width is increased by decays to
additional non-SM particles. Our main results are summarized in
Figs.~\ref{fig:xsec-G0}, \ref{fig:Rhigh} and \ref{fig:xsec-Gex} for
the 7 TeV LHC and Fig.~\ref{fig:4tat14} for the 14 TeV LHC. 
The large four-top cross sections found
in a large part of the parameter space, and their small SM backgrounds,
make  this 
channel a very promising probe of this class of models, capable to
reach gluon masses up to 800 GeV with the luminosity already collected
at the LHC.  An LHC energy upgrade to 8 (14) TeV implies an
  increase in the four-top production cross section by a factor of
  $\sim 2$
  (10-500), thus improving dramatically the reach in these models.

{\it Acknowledgements.} We would like to thank M. P\'erez-Victoria for
useful discussions and F. Veloso for help with statistical combination of limits.
This work has been partially supported by MICINN project
FPA2006-05294, FPA2010-17915
and Junta de Andaluc\'{\i}a projects FQM 101, FQM
03048 and FQM 6552. 

\end{document}